\documentclass{article}
\usepackage{amsmath}
\usepackage{amssymb}
\usepackage[dvips]{graphicx}
\begin{document}
\title{Newtonian and General Relativistic Models of Spherical Shells - II}
\author{D. Vogt\thanks{e-mail: dvogt@ime.unicamp.br} 
\and
P. S. Letelier\thanks{e-mail: letelier@ime.unicamp.br}\\
Departamento de Matem\'{a}tica Aplicada-IMECC, Universidade \\ 
Estadual de Campinas 13083-970 Campinas, S\~ao Paulo, Brazil}
\maketitle
\begin{abstract}
A family of potential-density pairs that represent spherical shells with finite 
thickness is obtained from the superposition of spheres with finite radii. 
Other families of shells with infinite thickness with a central hole are 
obtained by inversion transformations of spheres and of the finite shells. 
We also present a family of double shells with finite thickness. 
All potential-density pairs are analytical and can be stated in 
terms of elementary functions. For the 
above-mentioned structures, we study the circular orbits of test particles 
and their stability with respect to radial perturbations. All examples 
presented are found to be stable. A particular isotropic form of a metric 
in spherical coordinates is used to construct a General Relativistic version 
of the Newtonian families of spheres and shells. The matter of these 
structures is anisotropic, and the degree of anisotropy is a function of 
the radius.

\end{abstract}

\section{Introduction}
Analytical potential-density pairs with spherical symmetry have played an 
important role in galactic dynamics. Simplified models may be used not only to test 
and to perform sophisticated numerical simulations, but also to help us  
gain insight into more complicated phenomena. Several spherical analytical  
potential-density pairs have been proposed as models for elliptical 
galaxies and bulges of disc galaxies \cite{pl11,ja83,her90}, for 
dark matter haloes \cite{nav97}, as well as generalizations of these models
\cite{de93,tr94,zh96}. Shells constitute another spherical self-gravitating 
system that is important in issues like cosmology, gravitational collapse 
and supernovae (see, for instance, \cite{bs99} and references therein). 
A common simplification is to consider shells 
with infinitesimal thickness. Rein \cite{re99} constructed analytic solutions of the
Vlasov-Poisson and Vlasov-Einstein systems representing, respectively,
Newtonian and General Relativistic shells with finite thickness and 
a vacuum at the centre. A more detailed analysis of relativistic shells 
was done by \cite{an07a}. Solutions for relativistic double and also multishells
were found by \cite{an07b} using numerical methods. Other numerical 
solutions of finite Vlasov-Einstein shells were obtained by \cite{gr10}.
Recently, \cite{vl10} proposed a simple Newtonian
family of potential-density pairs of spherical shells with varying thicknesses and
also a General Relativistic version of these models.

In this work, we obtain families of potential-density pairs that
represent spherical shells with finite thickness and shells with infinite 
thickness that have a central hole. We also build a family of double shells 
with finite thickness. For these structures, we study the rotation curves 
(circular orbits) of test particles and make a first stability analysis 
by considering the stability of circular orbits under small radial perturbations.
We also consider a General Relativistic version of some of these Newtonian potential-density 
pairs in the same manner as in our previous work \cite{vl10}.  
The paper is divided as follows. In Section \ref{sec_sph}, we present a family 
of spheres with finite radii that will be used as a basis for the construction 
of the spherical shells. In Section \ref{sec_inv_sph}, the inversion theorem is 
applied on the spheres to obtain a first family of infinite shells. 
Then, in Section \ref{sec_sph_sh1}, shells with finite thickness are 
constructed by superposing different members of the family of spheres. 
The inversion theorem is also used on these shells to obtain another 
family of infinite shells with a central hole. In Section \ref{sec_sph_d_sh2},  
we show how to construct a family of double shells with finite 
thickness by superposing different members of finite shells, and we discuss 
a particular example. In Section \ref{sec_gr}, we use a  
particular isotropic form of a metric in spherical coordinates to construct 
a simple General Relativistic version of the Newtonian spheres and shells. We study in 
some detail two examples of spherical structures. Finally, in Section \ref{sec_dis} we summarize our 
results. 
 
\section{A family of finite spheres} \label{sec_sph}
We begin by considering a family of spheres with finite radius 
and following mass-density distribution,
\begin{equation} \label{eq_rho_sph1}
\rho_m=\begin{cases}
\rho_c \left(1-\frac{r^2}{a^2} \right)^{m-1/2}, & 0 \leq r \leq a,  \\
0, & r >a,
\end{cases}
\end{equation}
where $m=1,2,\dots$. This form of mass distribution has been chosen 
because it is a monotononically decreasing function of the radius and 
the corresponding potential can be expressed in terms of 
elementary functions, as will be seen below. Furthermore, models of thin finite discs 
with surface densities in the form of (\ref{eq_rho_sph1}), where $r$ now represents 
the cylindrical radius, are well known. The disc member with $m=1$ is the 
uniformly rotating disc of Kalnajs \cite{kal72}. Gonz\'alez \& Reina \cite{gon06} constructed a family of 
generalized Kalnajs discs with surface density similar to (\ref{eq_rho_sph1}); however,  
it is interesting to note that a family of discs with a similar density distribution has been used 
much earlier by Morgan \& Morgan \cite{mor69} in their study of General Relativistic static thin 
discs.  
  
The total mass $M_m$ of the sphere is
\begin{equation} \label{eq_m_sph} 
M_m=4 \pi \int_0^a r^2 \rho_m \mathrm{d}r=\frac{\rho_c\pi^2a^3(2m-1)!!}{2^m(m+1)!} \mbox{.}
\end{equation}
For convenience, we consider spheres with the same mass $M=M_m$. 
Then, equation (\ref{eq_rho_sph1}) can be rewritten as
\begin{equation} \label{eq_rho_sph2}
\rho_m=\frac{M2^m(m+1)!}{\pi^2a^3(2m-1)!!} \left(1-\frac{r^2}{a^2} \right)^{m-1/2} \mbox{.}
\end{equation}
The potential of the sphere with $m=1$ follows from direct integration of the
Poisson equation in spherical coordinates:
\begin{equation}
\frac{1}{r^2}\frac{\mathrm{d}}{\mathrm{d}r} \left( 
r^2 \frac{\mathrm{d}\Phi}{\mathrm{d}r} \right) =4\pi G \rho \mbox{,}
\end{equation} 
and is given by 
\begin{equation} \label{eq_phi_sph1}
\Phi_1 = \begin{cases}
-\frac{2GM}{3\pi a^4} \left[ \frac{3a^4}{r}\arcsin \left(
\frac{r}{a} \right) -\sqrt{a^2-r^2}
\left( 2r^2-5a^2 \right) \right], & 0 \leq r \leq a,  \\
-\frac{GM}{r}, & r>a.
\end{cases}
\end{equation}
The potentials of the spheres with $m>1$ can be found by using the
recurrence relation
\begin{equation}
\Phi_{m+1}=\frac{2m+4}{a^{2m+4}}\int a^{2m+3}\Phi_m \mathrm{d}a \mbox{.}
\end{equation}
For reference, the potentials of the members with $m=2,3,4$ read as
\begin{gather} 
\Phi_2= -\frac{2GM}{15\pi a^6} \left[ \frac{15a^6}{r} \arcsin \left( 
\frac{r}{a} \right) +\sqrt{a^2-r^2}
\left( 8r^4-26a^2r^2+33a^4 \right) \right] \mbox{,} \\
\Phi_3= -\frac{2GM}{105\pi a^8} \left[ \frac{105a^8}{r} \arcsin \left(
\frac{r}{a} \right)-\sqrt{a^2-r^2}
\left( 48r^6-200r^4a^2+326r^2a^4-279a^6 \right) \right] \mbox{,} \\
\Phi_4= -\frac{2GM}{315\pi a^{10}} \left[ \frac{315a^{10}}{r} \arcsin \left(
\frac{r}{a} \right)+\sqrt{a^2-r^2}
\left( 128r^8-656r^6a^2+1368r^4a^4 \right. \right. \notag \\
\left. \left. -1490r^2a^6+965a^8 \right) \right] \mbox{.}
\label{eq_phi_sph4}
\end{gather}

A common dynamical quantity used to characterize spherical distributions of
matter is the circular velocity $v_c$ (or the rotation curve) of test particles.
We will mosty be interested in the circular orbits of particles inside
the matter distributions. We also investigate the stability of these
orbits under small radial perturbations. This is characterized by the
epicyclic frequency $\kappa$, where stable circular orbits correspond to
$\kappa^2>0$. For spherical potentials, the circular velocity and 
epicyclic frequency are calculated through the relations \cite{bt08} 
\begin{equation}
v_c=\sqrt{r\Phi_{,r}} \text{,} \quad \text{and} \quad \kappa^2=\Phi_{,rr}+
\frac{3\Phi_{,r}}{r} \mbox{.}
\end{equation}
Expressions for the circular velocity $v_{c(m)}$ and epicyclic 
frequency $\kappa_m$ for the
spheres with $m=1,2,3,4$ are listed in Appendix \ref{ap_A}. Outside 
the spheres $(r>a)$, we have $v_c=(GM/r)^{1/2}$ and $\kappa=(GM/r^3)^{1/2}$.

\begin{figure}
\centering
\includegraphics[scale=0.65]{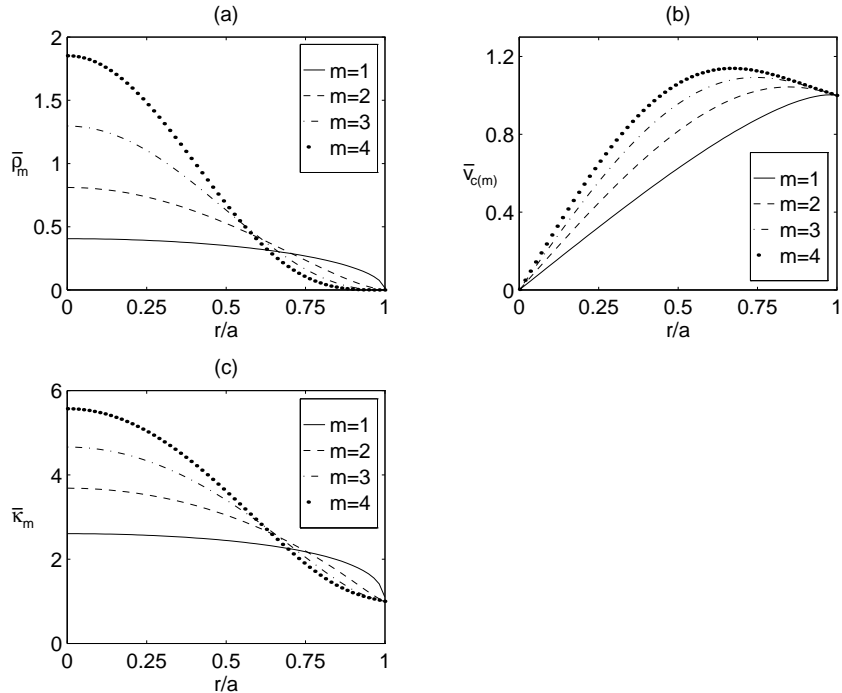}
\caption{(a) The mass density  $\bar{\rho}_m=\rho_m/(M/a^3)$, equation (\ref{eq_rho_sph2}), as
function of $r/a$, for four members of spheres. (b) The circular velocity 
$\bar{v}_{c(m)}=v_{c(m)}/(GM/a)^{1/2}$, equations (\ref{eq_vc1_sph})--(\ref{eq_vc4_sph}), 
and (c) the epicyclic frequency $\bar{\kappa}_m=\kappa_m/(GM/a^3)^{1/2}$, 
equations (\ref{eq_k1_sph})--(\ref{eq_k4_sph}), as functions of $r/a$, for the same 
members of spheres.} \label{fig1}
\end{figure}
In Fig.\ \ref{fig1}(a), we plot some curves of the dimensionless mass density 
$\bar{\rho}_m=\rho_m/(M/a^3)$,  
equation (\ref{eq_rho_sph2}), as function of $r/a$, for four members of the family of spheres.
In Figs \ref{fig1}(b) and (c) we show, respectively, some curves of the 
dimensionless circular velocity $\bar{v}_{c(m)}=v_{c(m)}/(GM/a)^{1/2}$, equations 
(\ref{eq_vc1_sph})--(\ref{eq_vc4_sph}), and
curves of the dimensionless epicyclic frequency $\bar{\kappa}_m=\kappa_m/(GM/a^3)^{1/2}$, 
equations (\ref{eq_k1_sph})--(\ref{eq_k4_sph}), 
as functions of $r/a$, for four members of the family of spheres. Unlike the disc case, we do not have 
a linear rotation profile for the sphere with $m=1$. We also note that the circular orbits
for the four first members are stable.

\section{Inverted spheres} \label{sec_inv_sph}
A new family of spherical potential density-pairs can be obtained by making 
an inversion (or a Kelvin transformation) \cite{kel47,kel53} on the spheres with
density (\ref{eq_rho_sph2}). In spherical coordinates, the inversion theorem 
states that  if a potential-density pair $\rho(r)$, $\Phi(r)$ is a solution
to the Poisson equation, then the pair 
\begin{gather}
\rho_{(inv.)}=\left( \frac{a}{r} \right)^5 \rho \left(\frac{a^2}{r} \right)
\mbox{,} \label{eq_sph_r_inv} \\
\Phi_{(inv.)}=\frac{a}{r} \Phi \left( \frac{a^2}{r} \right) \mbox{,} \label{eq_sph_p_inv}
\end{gather}
is also a solution to the Poisson equation. The inversion theorem applied to the density
(\ref{eq_rho_sph2}) results in 
\begin{equation} \label{eq_rho_inv1}
\rho_{m(inv.)}=\frac{M2^ma^2(m+1)!}{\pi^2r^5(2m-1)!!} \left(1-\frac{a^2}{r^2} \right)^{m-1/2}
\mbox{,} \quad r \geq a \mbox{.}
\end{equation}
This mass distribution can be interpreted as a family of spherical shells with 
a central hole of radius $a$. The shells extend to
infinity, although the density decays quite fast, $\rho_{m(inv.)} \propto 1/r^5$.
The maximum of density occurs at $r/a=\sqrt{(2m+4)/5}$. The mass $\mathcal{M}_m$
of each shell is
\begin{equation} 
\mathcal{M}_m=4 \pi \int_a^{\infty} r^2 \rho_{m(inv.)} \mathrm{d}r=
\frac{M2^{m+2}(m+1)!}{\pi (2m+1)(2m-1)!!} \mbox{.}
\end{equation}
For convenience, we consider shells with the same mass $\mathcal{M}=\mathcal{M}_m$ and
rewrite (\ref{eq_rho_inv1}) as
\begin{equation} \label{eq_rho_inv2}
\rho_{m(inv.)}=\frac{(2m+1)\mathcal{M}a^2}{4\pi r^5} \left(1-\frac{a^2}{r^2} \right)^{m-1/2}
\mbox{,} \quad r \geq a \mbox{.}
\end{equation}

The expressions for the potentials of the shells with $m=1,2,3,4$ are given
in Appendix \ref{ap_B}. From them result simple equations for the
rotation curves $v_{c(m)(inv.)}$ and epicyclic frequencies $\kappa_{m(inv.)}$, namely
\begin{gather} 
v_{c(1)(inv.)} =\sqrt{\frac{G\mathcal{M}}{r}} \left( 1-\frac{a^2}{r^2}\right)^{3/4} \mbox{,}
\quad v_{c(2)(inv.)} =\sqrt{\frac{G\mathcal{M}}{r}} \left( 1-\frac{a^2}{r^2}\right)^{5/4} 
\mbox{,} \label{eq_vc2_inv} \\
v_{c(3)(inv.)} =\sqrt{\frac{G\mathcal{M}}{r}} \left( 1-\frac{a^2}{r^2}\right)^{7/4} \mbox{,} 
\quad  v_{c(4)(inv.)} =\sqrt{\frac{G\mathcal{M}}{r}} \left( 1-\frac{a^2}{r^2}\right)^{9/4} 
\mbox{,} \label{eq_vc4_inv} \\
\kappa_{1(inv.)} =\sqrt{\frac{G\mathcal{M}(r^2+2a^2)}{r^5}} \left( 1-\frac{a^2}{r^2}
\right)^{1/4} \mbox{,} \label{eq_k1_inv} \\ 
\kappa_{2(inv.)} =\sqrt{\frac{G\mathcal{M}(r^2+4a^2)}{r^5}} \left( 1-\frac{a^2}{r^2}
\right)^{3/4} \mbox{,} \label{eq_k2_inv} \\
\kappa_{3(inv.)} =\sqrt{\frac{G\mathcal{M}(r^2+6a^2)}{r^5}} \left( 1-\frac{a^2}{r^2}
\right)^{5/4} \mbox{,} \label{eq_k3_inv} \\
\kappa_{4(inv.)} =\sqrt{\frac{G\mathcal{M}(r^2+8a^2)}{r^5}} \left( 1-\frac{a^2}{r^2}
\right)^{7/4} \mbox{,} \label{eq_k4_inv}
\end{gather}
which suggest the general relations
\begin{gather} 
v_{c(m)(inv.)}=\sqrt{\frac{G\mathcal{M}}{r}} \left( 1-\frac{a^2}{r^2}\right)^{(2m+1)/4} \mbox{,}  \\
\kappa_{m(inv.)}=\sqrt{\frac{G\mathcal{M} \left( r^2+2ma^2 \right)}{r^5}} \left( 1-\frac{a^2}{r^2}
\right)^{(2m-1)/4} \mbox{.} \label{eq_vk_inv} 
\end{gather}

\begin{figure}
\centering
\includegraphics[scale=0.65]{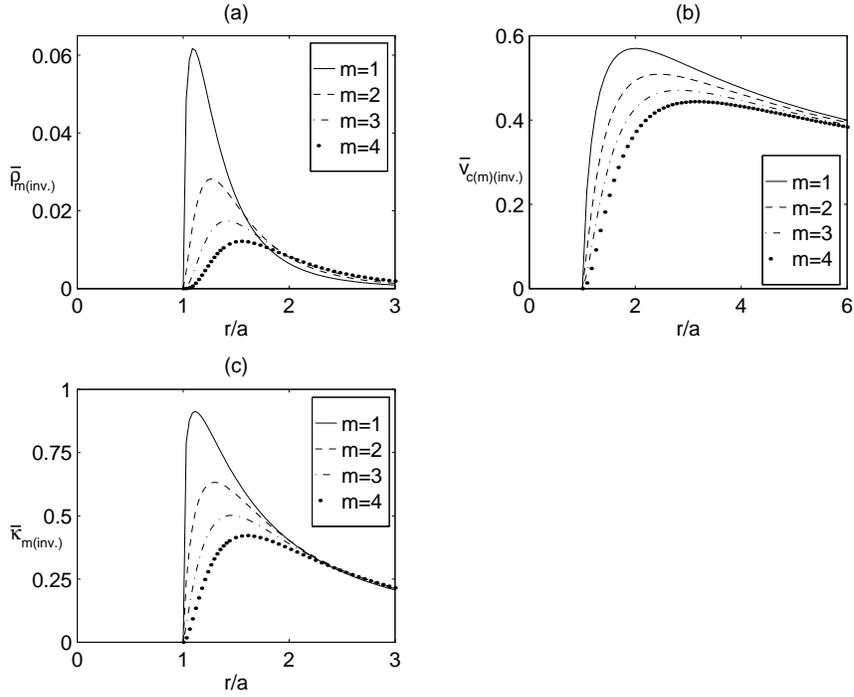}
\caption{(a) The mass density  $\bar{\rho}_{m(inv.)}=\rho_{m(inv.)}/(\mathcal{M}/a^3)$, equation (\ref{eq_rho_inv2}), as
function of $r/a$, for four members of inverted spheres. (b) The circular velocity 
$\bar{v}_{c(m)(inv.)}=v_{c(m)(inv.)}/(G\mathcal{M}/a)^{1/2}$, equations (\ref{eq_vc2_inv}) and (\ref{eq_vc4_inv}), 
and (c) the epicyclic frequency $\bar{\kappa}_{m(inv.)}=\kappa_{m(inv.)}/(G\mathcal{M}/a^3)^{1/2}$, 
equations (\ref{eq_k1_inv})--(\ref{eq_k4_inv}), as functions of $r/a$, for the same 
members of inverted spheres.} \label{fig2}
\end{figure}

Some curves of the dimensionless mass density $\bar{\rho}_{m(inv.)}=\rho_{m(inv.)}/(\mathcal{M}/a^3)$, 
equation (\ref{eq_rho_inv2}), as function of $r/a$,  are shown in Fig.\ \ref{fig2}(a). The shell with $m=1$ 
exibits a quite sharp density profile. In Figs \ref{fig2}(b) and (c), we plot, respectively, curves of the 
dimensionless circular velocity $\bar{v}_{c(m)(inv.)}=v_{c(m)(inv.)}/(G\mathcal{M}/a)^{1/2}$, equations 
(\ref{eq_vc2_inv}) and (\ref{eq_vc4_inv}), and the dimensionless epicyclic frequency 
$\bar{\kappa}_{m(inv.)}=\kappa_{m(inv.)}/(G\mathcal{M}/a^3)^{1/2}$, equations (\ref{eq_k1_inv})--(\ref{eq_k4_inv}), 
as functions of $r/a$. The four members of the shells shown are stable with respect to radial perturbations 
of circular orbits. From (\ref{eq_vk_inv}), we see that all members are
stable. According to Newton's shell
theorem, the potential inside the hole $(0 \leq r \leq a)$ is constant and thus the circular velocity
must be zero.  

\section{Families of spherical shells} \label{sec_sph_sh1}
Now we use the spheres discussed in Section \ref{sec_sph} to construct potential-density 
pairs that represent shells of matter with finite thickness. The method is similar to the used 
by \cite{let07} to obtain rings by superposing the Morgan \& Morgan discs \cite{mor69}. We take the 
density distribution (\ref{eq_rho_sph1}) and make the following sum: 
\begin{multline}  \label{eq_sup_sph}
\rho^{(m,n)}=\sum_{k=0}^{n} C^n_k(-1)^{n-k}\rho_{m+n+1-k} \\
=\rho_c \left( 1-\frac{r^2}{a^2} \right)^{m+1/2}\sum_{k=0}^{n} C^n_k(-1)^{n-k}
\left( 1-\frac{r^2}{a^2} \right)^{n-k}=
\rho_c \left( 1-\frac{r^2}{a^2} \right)^{m+1/2} \left( \frac{r}{a} \right)^{2n} \mbox{,}
\end{multline}
where $C^n_k=n!/[k!(n-k)!]$,  $m=0,1,\dots$ and $n=1,2,\dots$. This new density 
distribution vanishes on $r=0$ and $r=a$, so it represents a spherical shell with thickness 
$a$. Its mass density has a maximum value at $r/a=\sqrt{(2n)/(2m+2n+1)}$. The total mass 
$M^{(m,n)}$ of each shell is 
\begin{equation} 
M^{(m,n)}=4 \pi \int_0^a r^2 \rho^{(m,n)} \mathrm{d}r=
\frac{\rho_c\pi^2a^3(2m+1)!!(2n+1)!!}{2^{m+n+1}(m+n+2)!} \mbox{.}
\end{equation} 
The potentials associated with the mass density (\ref{eq_sup_sph}) are found by a similar 
sum, i.e. the potential of the shell with $m=0,n=1$ is $\Phi^{(0,1)}=\Phi_1-\Phi_2$. The 
explicit expressions for the potentials of the shells with $m=0,1$ and $n=1,2$ are given 
in Appendix \ref{ap_C}. 

\begin{figure}
\centering
\includegraphics[scale=0.65]{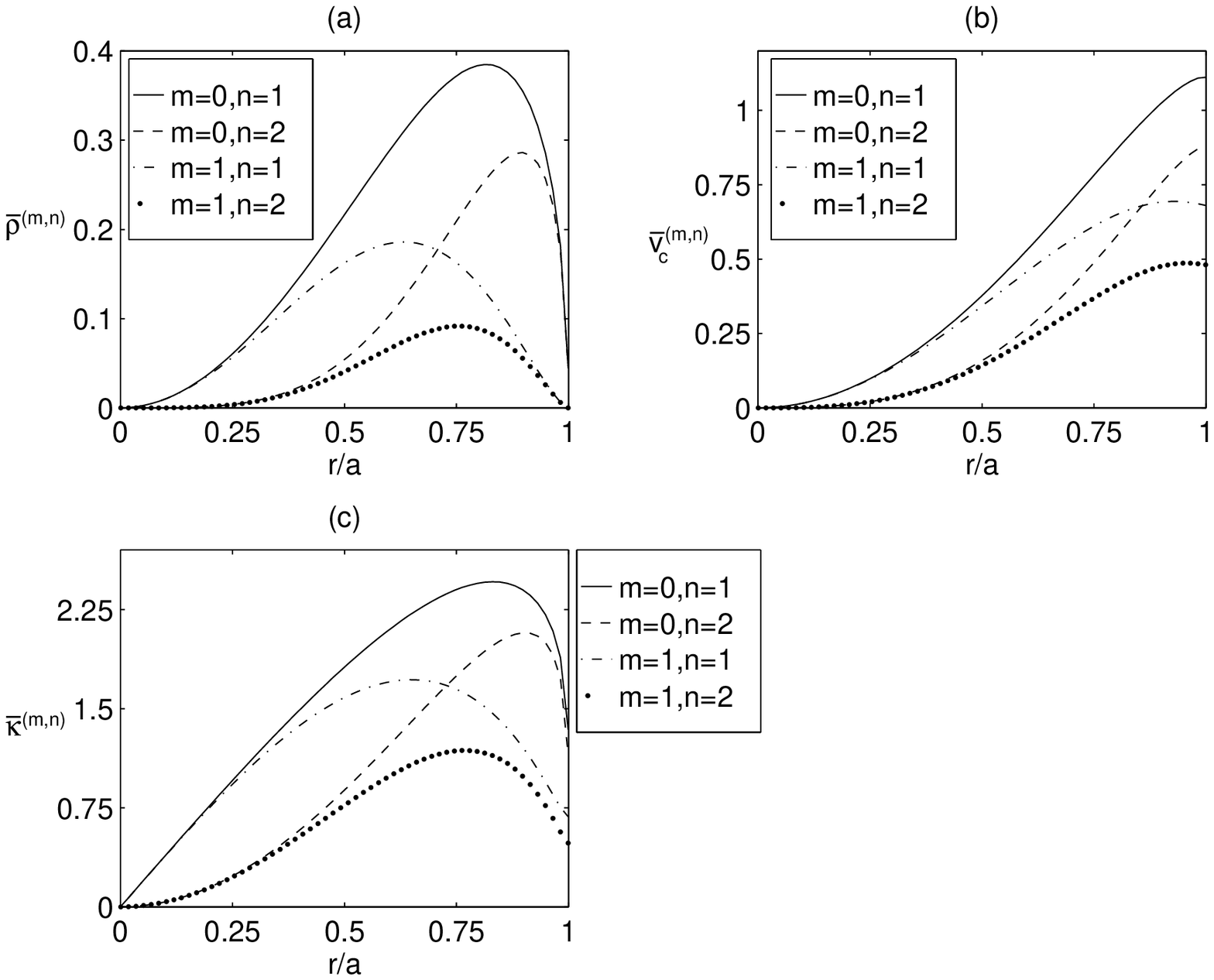}
\caption{(a) The mass density $\bar{\rho}^{(m,n)}=\rho^{(m,n)}/\rho_c$, equation (\ref{eq_sup_sph}), as
function of $r/a$, for four members of spherical shells. (b) The circular velocity
$\bar{v}_{c}^{(m,n)}=v_{c}^{(m,n)}/(G\rho_ca^2)^{1/2}$, equations (\ref{eq_vc_01})--(\ref{eq_vc_12}),
and (c) the epicyclic frequency $\bar{\kappa}^{(m,n)}=\kappa^{(m,n)}/(G\rho_c)^{1/2}$,
equations (\ref{eq_k_01})--(\ref{eq_k_12}), as functions of $r/a$, for the same
members of spherical shells.} \label{fig3}
\end{figure}

In Fig.\ \ref{fig3}(a), we display curves of the dimensionless mass density
$\bar{\rho}^{(m,n)}=\rho^{(m,n)}/\rho_c$, equation (\ref{eq_sup_sph}),
as function of $r/a$, for four members of spherical shells. 
For small values of $r/a$, the density of the
shell grows as $(r/a)^{2(n+1)}$; therefore, shells with larger values
of $n$ also have larger holes. Curves of the dimensionless circular velocity
$\bar{v}_{c}^{(m,n)}=v_{c}^{(m,n)}/(G\rho_ca^2)^{1/2}$,
equations (\ref{eq_vc_01})--(\ref{eq_vc_12}), and of the dimensionless
epicyclic frequency $\bar{\kappa}^{(m,n)}=\kappa^{(m,n)}/(G\rho_c)^{1/2}$,
equations (\ref{eq_k_01})--(\ref{eq_k_12}), are displayed in Figs \ref{fig3}(b)
and (c), respectively. For small values of $r/a$, the
rotation curves are proportional to $(r/a)^{n+2}$ and thus also grow
more slowly as $n$ increases. We also see that the members of shells shown in 
the figure are stable.

It is also possible to invert the shells. A Kelvin transformation applied to the density (\ref{eq_sup_sph}) 
results in 
\begin{equation} \label{eq_shell_inv}
\rho^{(m,n)}_{(inv.)}=\rho_c \left( 1-\frac{a^2}{r^2} \right)^{m+1/2}
\left( \frac{a}{r}\right)^{2n+5} \mbox{;} \quad r \geq a \mbox{,}
\end{equation}
which represents a shell with a central hole of radius $a$. If $n=0$ and $m\rightarrow m-1$ we 
recover the family of inverted shells discussed in Section \ref{sec_inv_sph}. They also extend 
to infinity with an even larger decay of density, $\rho^{(m,n)}_{(inv.)} \propto 1/r^{2n+5}$.  
The maximum of density occurs at $r/a=\sqrt{(2m+2n+6)/(2n+5)}$. The total mass
of each shell is
\begin{equation} 
4 \pi \int_a^{\infty} r^2 \rho^{(m,n)}_{(inv.)} \mathrm{d}r=
\frac{\rho_c\pi a^32^{n+2}n!(2m+1)!!}{(2m+2n+3)!!} \mbox{.}
\end{equation}

\begin{figure}
\centering
\includegraphics[scale=0.65]{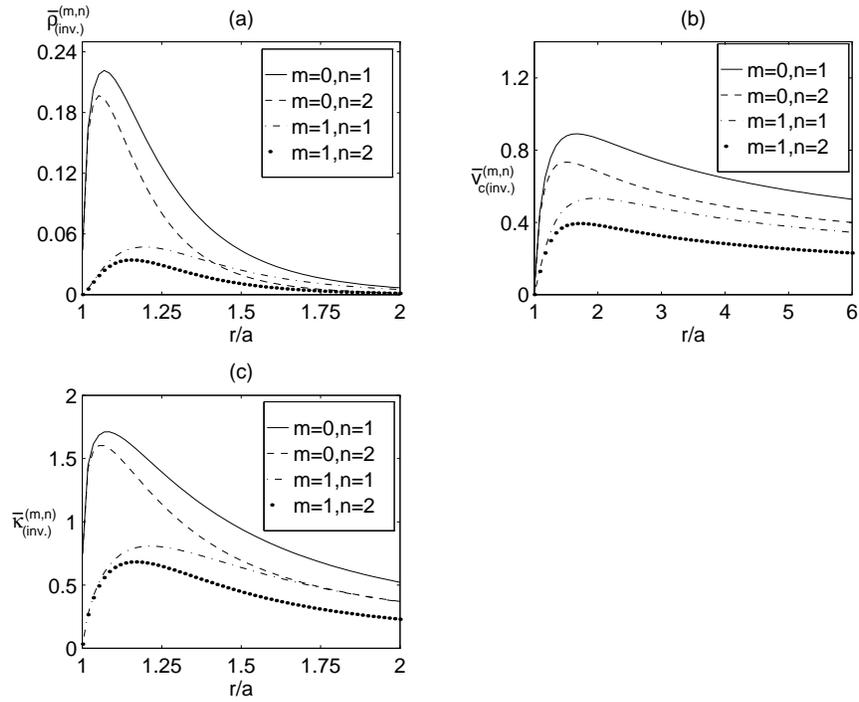}
\caption{(a) The mass density  $\bar{\rho}^{(m,n)}_{(inv.)}= \rho^{(m,n)}_{(inv.)}/\rho_c$, equation (\ref{eq_shell_inv}), as
function of $r/a$, for four members of inverted shells. (b) The circular velocity 
$\bar{v}^{(m,n)}_{c(inv.)}=v^{(m,n)}_{c(inv.)}/ (G\rho_ca^2)^{1/2}$, equations (\ref{eq_vc01_inv})--(\ref{eq_vc12_inv}),
and (c) the epicyclic frequency $\bar{\kappa}^{(m,n)}_{(inv.)}=\kappa^{(m,n)}_{(inv.)}/(G\rho_c)^{1/2}$, 
equations (\ref{eq_k01_inv})--(\ref{eq_k12_inv}), as functions of $r/a$, for the same 
members of inverted shells.} \label{fig4}
\end{figure}

Curves of the dimensionless mass density $\bar{\rho}^{(m,n)}_{(inv.)}$, equation (\ref{eq_shell_inv}), 
as function of $r/a$, are displayed in Fig.\ \ref{fig4}(a) for four members of inverted shells. 
Figs \ref{fig4}(b) and (c) show, respectively, curves of the dimensionless circular velocity
$\bar{v}^{(m,n)}_{c(inv.)}$, equations (\ref{eq_vc01_inv})--(\ref{eq_vc12_inv}), and of the dimensionless
epicyclic frequency $\bar{\kappa}^{(m,n)}_{(inv.)}$, equations (\ref{eq_k01_inv})--(\ref{eq_k12_inv}), 
for the same members of inverted shells. Once again, these members are stable with respect
to radial perturbations of circular orbits.

\section{A family of double shells} \label{sec_sph_d_sh2}

The members of the family of shells discussed in Section \ref{sec_sph_sh1} may be 
combined to generate double shells with finite
thickness. To achieve this, we superpose members of (\ref{eq_sup_sph}) in the 
following way:
\begin{equation} \label{eq_sup_double}
\rho^{(m,n)}_{(2)}=\rho^{(m,n)}-2b^2\rho^{(m,n+1)}+b^4\rho^{(m,n+2)} 
=\rho_c \left( 1-\frac{r^2}{a^2} \right)^{m+1/2} \left( \frac{r}{a} \right)^{2n} 
\left( 1-\frac{b^2r^2}{a^2} \right)^2 \mbox{,}
\end{equation}
where $b>1$. This density distribution may be interpreted as 
two concentric spherical shells with a gap located at $r/a=1/b$. 
We take as an example the member with $m=0,n=1$. The potential as well as
the expressions of the rotation curve and epicyclic frequency 
for this example are given in Appendix \ref{ap_E}. 

\begin{figure}
\centering
\includegraphics[scale=0.65]{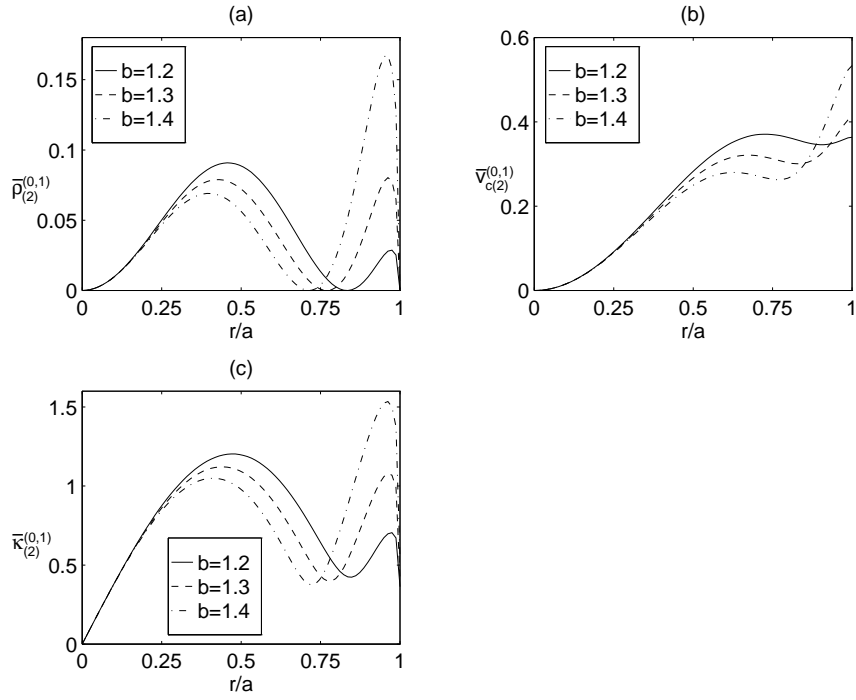}
\caption{(a) The mass density $\bar{\rho}^{(0,1)}_{(2)}=\rho^{(0,1)}_{(2)}/\rho_c$, equation (\ref{eq_sup_double}), as
function of $r/a$, for a double shell. (b) The circular velocity
$\bar{v}_{c(2)}^{(0,1)}=v_{c(2)}^{(0,1)}/(G\rho_ca^2)^{1/2}$, equation (\ref{eq_vc_2_01}),
and (c) the epicyclic frequency $\bar{\kappa}_{(2)}^{(m,n)}=\kappa_{(2)}^{(m,n)}/(G\rho_c)^{1/2}$,
equation (\ref{eq_k_2_01}), as functions of $r/a$, for the same double shell.} \label{fig5}
\end{figure}

Fig.\ \ref{fig5}(a) shows curves of the dimensionless mass density
$\bar{\rho}^{(0,1)}_{(2)}=\rho^{(0,1)}_{(2)}/\rho_c$, equation (\ref{eq_sup_double}), as 
function of $r/a$ for some values of $b$. Curves of the circular velocity
$\bar{v}_{c(2)}^{(0,1)}=v_{c(2)}^{(0,1)}/(G\rho_ca^2)^{1/2}$, equation (\ref{eq_vc_2_01}), 
and of the epicyclic frequency $\bar{\kappa}_{(2)}^{(m,n)}=\kappa_{(2)}^{(m,n)}/(G\rho_c)^{1/2}$,
equation (\ref{eq_k_2_01}), are plotted in Figs \ref{fig5}(b) and (c),
respectively. Unlike the structures discussed so far, the rotation curves for this 
example of double shell show a point of minimum. They also show no sign of instability.

\section{General Relativistic Spheres and Shells} \label{sec_gr}
Now we consider an extension to General Relativity of the Newtonian 
spherical potential-density
pairs discussed so far. We choose a particular metric
in an isotropic form in spherical coordinates $(t,r,\theta,\varphi)$, which was also
used in a recent work on another relativistic model of shells \cite{vl10},
\begin{equation} \label{eq_metric}
\mathrm{d}s^2=\left( \frac{1-f}{1+f} \right)^2c^2\mathrm{d}t^2-\left( 1+f \right)^4 
\left( \mathrm{d}r^2+r^2\mathrm{d}\theta^2
+r^2\sin^2 \theta \mathrm{d} \varphi^2 \right) \mbox{,}
\end{equation}
where $f=f(r)$. The Schwarzschild solution is given by metric (\ref{eq_metric}) if 
$f=GM/(2c^2r)$, and the relation between the function $f(r)$ and the Newtonian 
potential $\Phi(r)$ is 
\begin{equation} \label{eq_f_phi}
f=-\frac{\Phi}{2c^2} \mbox{.} 
\end{equation}
By using the Einstein equations, we find the following expressions for the non-zero
components of the energy-momentum tensor $T_{\mu\nu}$ \cite{vl10}
\begin{gather}
T^t_t=-\frac{c^4}{2\pi G \left( 1+f \right)^5} \frac{1}{r^2} 
\frac{\mathrm{d}}{\mathrm{d}r} \left( r^2\frac{\mathrm{d}f}{\mathrm{d}r} \right) \mbox{,} \label{eq_Ttt} \\
T^r_r=\frac{c^4}{2\pi G \left( 1+f \right)^5\left( 1-f \right)} \frac{\mathrm{d}f}{\mathrm{d}r} 
\left( \frac{f}{r}+\frac{\mathrm{d}f}{\mathrm{d}r} \right) \mbox{,} \\
T^{\theta}_{\theta}=T^{\varphi}_{\varphi}=\frac{c^4}{4\pi G \left( 1+f \right)^5\left( 1-f \right)} 
\left[ f\frac{\mathrm{d}^2f}{\mathrm{d}r^2}+\frac{f}{r}\frac{\mathrm{d}f}{\mathrm{d}r}
-\left( \frac{\mathrm{d}f}{\mathrm{d}r} \right)^2 \right] \mbox{.} \label{eq_Tthth}
\end{gather}
The energy density reads as $\epsilon=T^t_t/c^2$ and the pressures or tensions
along a direction $k$ are given by $P_k=-T^k_k$. The `effective Newtonian density'
$\rho_N=\epsilon+P_r/c^2+P_{\theta}/c^2+P_{\varphi}/c^2$ can be cast as 
\begin{equation} \label{eq_rho_N1}
\rho_N=-\frac{c^2}{2\pi G \left( 1+f \right)^5\left( 1-f \right)} \frac{1}{r^2}
\frac{\mathrm{d}}{\mathrm{d}r} \left( r^2\frac{\mathrm{d}f}{\mathrm{d}r} \right) \mbox{.}
\end{equation}

We will study in some detail two simple examples of General Relativistic spherical
structures with finite extent: the sphere with potential (\ref{eq_phi_sph1}) and
the shell with potential (\ref{eq_phi_01}). Using equations (\ref{eq_phi_sph1}) and
(\ref{eq_f_phi})--(\ref{eq_rho_N1}), we find the following expressions for the 
components of the energy-momentum tensor of the sphere,
\begin{gather}
\tilde{\epsilon}_1=\frac{4}{\pi^2\tilde{a}^3\left( 1+f_1 \right)^5}
\sqrt{1-\frac{\tilde{r}^2}{\tilde{a}^2}} \mbox{,} \label{eq_eps_gr1} \\
\tilde{\rho}_{N1}=\frac{4}{\pi^2\tilde{a}^3\left( 1+f_1 \right)^5\left( 1-f_1 \right)}
\sqrt{1-\frac{\tilde{r}^2}{\tilde{a}^2}} \mbox{,} \label{eq_rhon_gr1} \\
\tilde{P}_{r1}=\frac{8}{3\pi^3\tilde{a}^4\tilde{r}^2\left( 1+f_1 \right)^5\left( 1-f_1 \right)}
\left( 1-\frac{\tilde{r}^2}{\tilde{a}^2} \right)^{3/2} \left[ \frac{\tilde{a}^3}{\tilde{r}}
\arcsin \left( \frac{\tilde{r}}{\tilde{a}} \right) \right. \notag \\
\left. +\left( 2\tilde{r}^2-\tilde{a}^2\right)
\sqrt{1-\frac{\tilde{r}^2}{\tilde{a}^2}} \right] \mbox{,} \label{eq_pr_gr1} \\
\tilde{P}_{\theta 1}=\tilde{P}_{\varphi 1}=\frac{4}{3\pi^3\tilde{a}^4\tilde{r}^2
\left( 1+f_1 \right)^5\left( 1-f_1 \right)} \sqrt{1-\frac{\tilde{r}^2}{\tilde{a}^2}}
\left[ \frac{\tilde{a}}{\tilde{r}} \left( 4\tilde{r}^2-\tilde{a}^2\right)
\arcsin \left( \frac{\tilde{r}}{\tilde{a}} \right) \right. \notag \\
\left. + \left( 2\tilde{r}^2+\tilde{a}^2\right)
\sqrt{1-\frac{\tilde{r}^2}{\tilde{a}^2}} \right] \label{eq_pth_gr1} \mbox{,}
\end{gather}
with
\begin{equation}
f_1=\frac{2}{3\pi \tilde{a}^4} \left[ \frac{3\tilde{a}^4}{\tilde{r}}\arcsin \left(
\frac{\tilde{r}}{\tilde{a}} \right) -\sqrt{\tilde{a}^2-\tilde{r}^2}
\left( 2\tilde{r}^2-5\tilde{a}^2 \right) \right] \mbox{,}
\end{equation}
and the dimensionless variables and parameters are $\tilde{r}=r/r_s$, $\tilde{a}=a/r_s$,
$\tilde{\epsilon}_1=\epsilon_1/(M/r_s^3)$, $\tilde{\rho}_{N1}=\rho_{N1}/(M/r_s^3)$,
$\tilde{P}_{k1}=P_{k1}/(Mc^2/r_s^3)$ and $r_s=GM/(2c^2)$.

\begin{figure}
\centering
\includegraphics[scale=0.65]{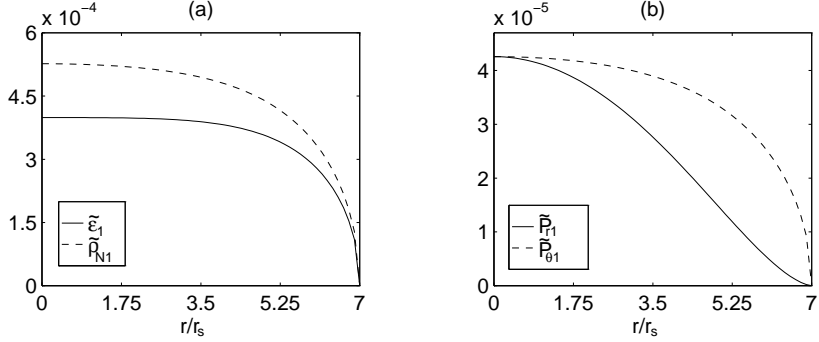}
\caption{(a) The energy density $\tilde{\epsilon}_1=\epsilon_1/(M/r_s^3)$ (solid curve),
equation (\ref{eq_eps_gr1}), and the effective Newtonian density
$\tilde{\rho}_{N1}=\rho_{N1}/(M/r_s^3)$ (dashed curve), equation (\ref{eq_rhon_gr1}), for the sphere
with radius $\tilde{a}=7$. (b) The radial pressure $\tilde{P}_{r1}=P_{r1}/(Mc^2/r_s^3)$ (solid curve),
equation (\ref{eq_pr_gr1}), and the azimuthal/polar pressure 
$\tilde{P}_{\theta 1}=P_{\theta 1}/(Mc^2/r_s^3)$ (dashed curve), equation (\ref{eq_pth_gr1}), 
for the sphere with radius $\tilde{a}=7$.} \label{fig6}
\end{figure}

In order to represent physically meaningful matter distributions, the components of the energy-momentum
tensor should satisfy the energy conditions. The strong energy condition states that 
$\rho_N \geq 0$, whereas the weak energy condition imposes the condition $\epsilon \geq 0$. 
The dominant energy condition requires $|P_k/\epsilon| \leq c^2$. Equations 
(\ref{eq_eps_gr1}) and (\ref{eq_rhon_gr1}) show that the strong and weak energy
conditions are satisfied if $\tilde{a} \geq 16/(3\pi) \approx 1.70$. In Fig.\ 
\ref{fig6}(a), we plot the curves of the energy density $\tilde{\epsilon}_1$ (solid
curve) and the effective Newtonian density $\tilde{\rho}_{N1}$ (dashed curve),
as functions of $\tilde{r}=r/r_s$, for the sphere with radius $\tilde{a}=7$,
and in Fig.\ \ref{fig6}(b), we display the radial pressure $\tilde{P}_{r1}$
(solid curve) and the azimuthal/polar pressure $\tilde{P}_{\theta 1}$, 
as functions of $\tilde{r}=r/r_s$, for the same sphere. Unfortunately, we
found that the energy density is not always a monotone decreasing function 
of the radius. This happens if $\tilde{a} \lessapprox 6.8$. For $\tilde{a} \gtrapprox 6.8$,
the dominant energy condition is also satisfied. The radial and azimuthal
pressures also decrease monotonically and both vanish on the sphere's surface.
In this example, we have an anisotropic sphere with larger azimuthal/polar than
radial pressures.

In the case of the shell with potential (\ref{eq_phi_01}), we obtain the following
expressions for the components of the energy-momentum tensor:
\begin{gather} 
\hat{\epsilon}^{(0,1)}=\frac{1}{\left[ 1+f^{(0,1)} \right]^5}\left( \frac{\hat{r}}
{\hat{a}}\right)^2 \sqrt{1-\frac{\hat{r}^2}{\hat{a}^2}} \mbox{,} \label{eq_eps_gr2} \\
\tilde{\rho}_{N}^{(0,1)}=\frac{1}{\left[ 1+f^{(0,1)} \right]^5\left[ 1-f^{(0,1)} \right]}
\left( \frac{\hat{r}}{\hat{a}}\right)^2 \sqrt{1-\frac{\hat{r}^2}{\hat{a}^2}} \mbox{,} \label{eq_rhon_gr2} \\
\hat{P}_{r}^{(0,1)}=\frac{\pi\left( 3\hat{r}^2+2\hat{a}^2\right)}
{360\tilde{a}^2\tilde{r}^2\left[ 1+f^{(0,1)} \right]^5\left[ 1-f^{(0,1)} \right]}
\left( 1-\frac{\hat{r}^2}{\hat{a}^2} \right)^{3/2} \left[ \frac{3\hat{a}^5}{\hat{r}}
\arcsin \left( \frac{\hat{r}}{\hat{a}} \right) \right. \notag \\
\left. +\left( 2\hat{r}^2+\hat{a}^2\right)
\left( 4\hat{r}^2-3\hat{a}^2\right)\sqrt{1-\frac{\hat{r}^2}{\hat{a}^2}} \right] \mbox{,} \label{eq_pr_gr2} \\
\hat{P}_{\theta}^{(0,1)}=\hat{P}_{\varphi}^{(0,1)}=\frac{\pi}
{720\tilde{a}^2\tilde{r}^2\left[ 1+f^{(0,1)} \right]^5\left[ 1-f^{(0,1)} \right]}
\sqrt{1-\frac{\hat{r}^2}{\hat{a}^2}}  \notag \\
\times \left[ \frac{3\hat{a}^3}{\hat{r}}
\left( 18\hat{r}^4-\hat{r}^2\hat{a}^2-2\hat{a}^4\right)
\arcsin \left( \frac{\hat{r}}{\hat{a}} \right) 
+\left( 4\hat{r}^6+28\hat{r}^4\hat{a}^2
+7\hat{r}^2\hat{a}^4+6\hat{a}^6\right)
\sqrt{1-\frac{\hat{r}^2}{\hat{a}^2}} \right] \label{eq_pth_gr2} \mbox{,}
\end{gather}
with 
\begin{equation}
f^{(0,1)}=\frac{1}{120\hat{a}^3} \left[ \frac{15\hat{a}^6}{\hat{r}} 
\arcsin \left( \frac{\hat{r}}{\hat{a}} \right) -
\sqrt{\hat{a}^2-\hat{r}^2}\left( 8\hat{r}^4-6\hat{a}^2\hat{r}^2-17\hat{a}^4\right) \right] \mbox{,}
\end{equation}
and the dimensionless variables and parameters are now $\hat{r}=r/r_c$, $\hat{a}=a/r_c$,
$\hat{\epsilon}^{(0,1)}=\epsilon^{(0,1)}/\rho_c$, 
$\hat{\rho}_{N}^{(0,1)}=\rho_{N}^{(0,1)}/\rho_c$,
$\hat{P}_{k}^{(0,1)}=P_{k}^{(0,1)}/(\rho_cc^2)$ and $r_c=c/\sqrt{G\rho_c}$.

\begin{figure}
\centering
\includegraphics[scale=0.65]{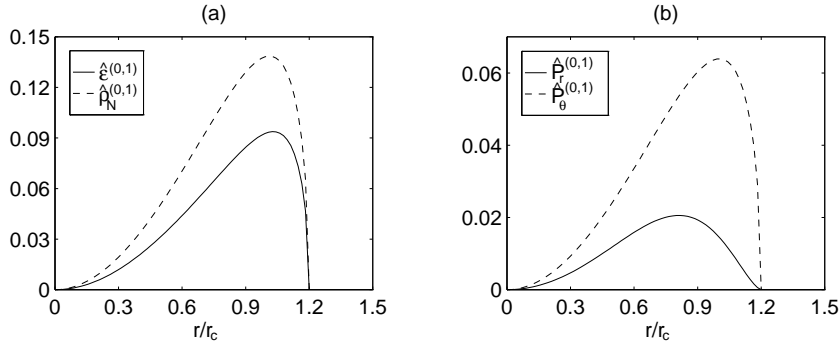}
\caption{(a) The energy density $\hat{\epsilon}^{(0,1)}=\epsilon^{(0,1)}/\rho_c$ (solid curve),
equation (\ref{eq_eps_gr2}), and the effective Newtonian density
$\hat{\rho}_{N}^{(0,1)}=\rho_{N}^{(0,1)}/\rho_c$ (dashed curve), equation (\ref{eq_rhon_gr2}), for the shell
with $\hat{a}=1.2$. (b) The radial pressure $\hat{P}_{r}^{(0,1)}=P_{r}^{(0,1)}/(\rho_cc^2)$ (solid curve),
equation (\ref{eq_pr_gr2}), and the azimuthal/polar pressure
$\hat{P}_{\theta}^{(0,1)}=P_{\theta}^{(0,1)}/(\rho_cc^2)$ (dashed curve), equation (\ref{eq_pth_gr2}),
for the shell with $\hat{a}=1.2$.} \label{fig7}
\end{figure}

From equations (\ref{eq_eps_gr2}) and (\ref{eq_rhon_gr2}), we have that the strong and
weak energy conditions are satisfied when $\hat{a} \leq \sqrt{15}/2 \approx 1.94$. 
The dominant energy condition holds if 
\begin{equation}
\hat{a} \leq \frac{5\sqrt{6\pi +15}}{2(2\pi +5)} \approx 1.29 \mbox{.} 
\end{equation}
Fig.\ \ref{fig7}(a) shows curves of the energy density $\hat{\epsilon}^{(0,1)}$ (solid curve) 
and  the effective Newtonian density $\hat{\rho}_{N}^{(0,1)}$ (dashed curve), as functions 
of  $r/r_c$, for the shell with parameter $\hat{a}=1.2$. In Fig.\ \ref{fig7}(b), we show curves
of the radial pressure $\hat{P}_{r}^{(0,1)}$ (solid curve) and the azimuthal/polar pressure
$\hat{P}_{\theta}^{(0,1)}$ (dashed curve), as functions of $r/r_c$, for the same shell. 
The pressures, as well as the densities, vanish at the centre and and at the outer surface of 
the shell. This example shows a highly anisotropic relativistic shell with larger azimuthal/polar than
radial pressures. We note that the analytical and numerical solutions for relativistic finite shells
found by \cite{an07a} and \cite{gr10} also have anisotropic pressures.
\section{Discussion} \label{sec_dis}
In this work, we constructed families of potential-density pairs that represent
spherical shells with finite thickness and shells with infinite thickness, but 
with a hole at the centre. Shells with finite thickness
were obtained by the superposition of spheres with finite radii, and 
shells with infinite extend by using the inversion theorem on spheres and finite
shells. Furthermore, shells with finite thickness may be superposed to give
a family of double shells with finite thickness. All the
potential-density pairs discussed in this work can be expressed in terms of elementary functions. 
We also studied the rotation curves for these self-gravitating structures and made a stability
analysis based on radial perturbations of circular orbits of individual particles.
In all the examples discussed, we found stable structures.

A General Relativistic version of the Newtonian spherical potential-density pairs 
was obtained by using a particular isotropic form of a metric in spherical coordinates. 
We studied in some detail an example of a relativistic sphere and of a shell. In both 
cases, the matter presents equal azimuthal and polar pressures, but different from
the radial pressure, and this anisotropy is a function of radius. It is possible to 
choose parameters such that all energy conditions are satisfied.  
\section*{Acknowledgments}

D.\ V.\ thanks FAPESP for financial support, P.\ S.\ L.\ thanks FAPESP and CNPq
for partial financial support.

\appendix
\section{Rotation curves and epicyclic frequency for spheres} \label{ap_A}

The expressions for the circular velocity $v_{c(m)}$ and epicyclic frequency
$\kappa_m$ for the spheres with $m=1,2,3,4$ read as
\begin{gather} 
v_{c(1)}=\sqrt{\frac{2GM}{\pi a^4}} \left[ \frac{a^4}{r} \arcsin \left( 
\frac{r}{a} \right)+\sqrt{a^2-r^2} \left( 2r^2-a^2 \right) \right]^{1/2} 
\mbox{,} \label{eq_vc1_sph}\\
v_{c(2)}=\sqrt{\frac{2GM}{3\pi a^6}} \left[ \frac{3a^6}{r} \arcsin \left( 
\frac{r}{a} \right) - \sqrt{a^2-r^2} \left( 8r^4-14r^2a^2+3a^4 \right)
 \right]^{1/2} \mbox{,} \\
v_{c(3)}=\sqrt{\frac{2GM}{15\pi a^8}} \left[ \frac{15a^8}{r} \arcsin \left(
\frac{r}{a} \right)+ \sqrt{a^2-r^2} \left( 48r^6-136r^4a^2+118r^2a^4-15a^6 \right)
\right]^{1/2} \mbox{,} \\
v_{c(4)}=\sqrt{\frac{2GM}{105\pi a^{10}}} \left[ \frac{105a^{10}}{r} \arcsin \left(
\frac{r}{a} \right)- \sqrt{a^2-r^2} \left( 384r^8-1488r^6a^2 \right. \right. \notag \\
\left. \left. +2104r^4a^4-1210r^2a^6+105a^8 \right)
\right]^{1/2} \mbox{,} \label{eq_vc4_sph}\\
\kappa_1= \sqrt{\frac{2GM}{\pi r^2a^4}} \left[ \frac{a^4}{r} \arcsin \left(
\frac{r}{a} \right)+\sqrt{a^2-r^2} \left( 10r^2-a^2 \right)
\right]^{1/2} \mbox{,} \label{eq_k1_sph} \\
\kappa_2=\sqrt{\frac{2GM}{3\pi r^2a^6}} \left[ \frac{3a^6}{r} \arcsin \left(
\frac{r}{a} \right)-\sqrt{a^2-r^2} \left( 56r^4-62r^2a^2+3a^4 \right)
\right]^{1/2} \mbox{,} \\
\kappa_3=\sqrt{\frac{2GM}{15\pi r^2a^8}} \left[ \frac{15a^8}{r} \arcsin \left(
\frac{r}{a} \right)+\sqrt{a^2-r^2} \left( 432r^6-904r^4a^2+502r^2a^4-15a^6 \right)
\right]^{1/2} \mbox{,} \\
\kappa_4=\sqrt{\frac{2GM}{105\pi r^2a^{10}}} \left[ \frac{105a^{10}}{r} \arcsin \left(
\frac{r}{a} \right)-\sqrt{a^2-r^2} \left( 4224r^8-13008r^6a^2 \right. \right. \notag \\
\left. \left. +13624r^4a^4-5050r^2a^6+105a^8 \right)
\right]^{1/2} \mbox{.} \label{eq_k4_sph}
\end{gather}

\section{Potentials of inverted spheres} \label{ap_B}
The expressions for the potentials $\Phi_{m(inv.)}$ of the shells (\ref{eq_rho_inv2}) with $m=1,2,3,4$, obtained by
inversion of spheres, are
\begin{gather}
\Phi_{1(inv.)}=-\frac{G\mathcal{M}}{8r}\left[ \frac{3r}{a} \arcsin \left( \frac{a}{r} \right) + 
\frac{\sqrt{r^2-a^2}}{r^3} \left( 5r^2-2a^2 \right) \right] \mbox{,} \\
\Phi_{2(inv.)}=-\frac{G\mathcal{M}}{48r}\left[ \frac{15r}{a} \arcsin \left( \frac{a}{r} \right) + 
\frac{\sqrt{r^2-a^2}}{r^5} \left( 33r^4-26r^2a^2+8a^4 \right) \right] \mbox{,} \\
\Phi_{3(inv.)}=-\frac{G\mathcal{M}}{384r}\left[ \frac{105r}{a}  \arcsin \left( \frac{a}{r} \right) +
\frac{\sqrt{r^2-a^2}}{r^7} \left( 279r^6-326r^4a^2+200r^2a^4-48a^6 \right) \right] \mbox{,} \\
\Phi_{4(inv.)}=-\frac{G\mathcal{M}}{1280r}\left[ \frac{315r}{a}  \arcsin \left( \frac{a}{r} \right) +
\frac{\sqrt{r^2-a^2}}{r^9} \left( 965r^8-1490r^6a^2 \right. \right. \notag \\
\left. \left. +1368r^4a^4-656r^2a^6+128a^8 \right) \right] \mbox{.} 
\end{gather}

\section{Potentials, rotation curves and epicyclic frequencies for spherical shells} \label{ap_C}
The expressions for the potentials $\Phi^{(m,n)}$ of the shells (\ref{eq_sup_sph}) with $m=0,1$ 
and $n=1,2$ are 
\begin{gather}
\Phi^{(0,1)}=-\frac{G\rho_c\pi}{60a^3} \left[ \frac{15a^6}{r} \arcsin \left( \frac{r}{a} \right) -
\sqrt{a^2-r^2}\left( 8r^4-6a^2r^2-17a^4\right) \right] \mbox{,} \label{eq_phi_01} \\
\Phi^{(0,2)}=-\frac{G\rho_c\pi}{3360a^5} \left[ \frac{525a^8}{r} \arcsin \left( \frac{r}{a} \right) -
\sqrt{a^2-r^2}\left( 240r^6-104r^4a^2 \right. \right. \notag \\
\left. \left. -162r^2a^4-499a^6\right) \right] \mbox{,} \\
\Phi^{(1,1)}=-\frac{G\rho_c\pi}{1120a^5} \left[ \frac{105a^8}{r} \arcsin \left( \frac{r}{a} \right)+
\sqrt{a^2-r^2}\left( 80r^6-184r^4a^2 \right. \right. \notag \\
\left. \left. +58r^2a^4+151a^6\right) \right] \mbox{,} \\
\Phi^{(1,2)}=-\frac{G\rho_c\pi}{20160a^7} \left[ \frac{945a^{10}}{r} \arcsin \left( \frac{r}{a} \right)+
\sqrt{a^2-r^2}\left( 896r^8-1712r^6a^2 \right. \right. \notag \\
\left. \left. +264r^4a^4+394r^2a^6+1103a^8\right) \right] \mbox{.} 
\label{eq_phi_12}
\end{gather}
We also give the expressions for the circular velocity $v_c^{(m,n)}$ and 
epicyclic frequency $\kappa^{(m,n)}$ for these members:
\begin{gather} 
v_c^{(0,1)}=\sqrt{\frac{G\rho_c\pi}{12a^3}}\left[ \frac{3a^6}{r} \arcsin \left( \frac{r}{a} \right)
+\sqrt{a^2-r^2} \left( 8r^4-2r^2a^2-3a^4 \right) \right]^{1/2} \mbox{,} \label{eq_vc_01} \\
v_c^{(0,2)}=\sqrt{\frac{G\rho_c\pi}{96a^5}}\left[ \frac{15a^8}{r} \arcsin \left( \frac{r}{a} \right) +
\sqrt{a^2-r^2} \left( 48r^6-8r^4a^2-10r^2a^4-15a^6 \right) \right]^{1/2} \mbox{,} \\
v_c^{(1,1)}=\sqrt{\frac{G\rho_c\pi}{32a^5}}\left[ \frac{3a^8}{r} \arcsin \left( \frac{r}{a} \right) - 
\sqrt{a^2-r^2} \left( 16r^6-24r^4a^2+2r^2a^4+3a^6 \right) \right]^{1/2} \mbox{,} \\
v_c^{(1,2)}=\sqrt{\frac{G\rho_c\pi}{320a^7}}\left[ \frac{15a^{10}}{r} \arcsin \left( \frac{r}{a} \right) -
\sqrt{a^2-r^2} \left( 128r^8-176r^6a^2+8r^4a^4 \right. \right. \notag \\
\left. \left. +10r^2a^6+15a^8 \right) \right]^{1/2} \mbox{,} \label{eq_vc_12}  \\
\kappa^{(0,1)}= \sqrt{\frac{G\rho_c\pi}{12r^2a^3}}\left[ \frac{3a^6}{r} \arcsin \left( \frac{r}{a} \right)
+\sqrt{a^2-r^2} \left( 56r^4-2r^2a^2-3a^4 \right) \right]^{1/2} \mbox{,} \label{eq_k_01} \\
\kappa^{(0,2)}=\sqrt{\frac{G\rho_c\pi}{96r^2a^5}}\left[ \frac{15a^8}{r} \arcsin \left( \frac{r}{a} \right) +
\sqrt{a^2-r^2} \left( 432r^6-8r^4a^2-10r^2a^4-15a^6 \right) \right]^{1/2} \mbox{,} \\
\kappa^{(1,1)}=\sqrt{\frac{G\rho_c\pi}{32r^2a^5}}\left[ \frac{3a^8}{r} \arcsin \left( \frac{r}{a} \right) - 
\sqrt{a^2-r^2} \left( 144r^6-152r^4a^2+2r^2a^4+3a^6 \right) \right]^{1/2} \mbox{,} \\
\kappa^{(1,2)}=\sqrt{\frac{G\rho_c\pi}{320r^2a^7}}\left[ \frac{15a^{10}}{r} \arcsin \left( \frac{r}{a} \right) -
\sqrt{a^2-r^2} \left( 1408r^8-1456r^6a^2 \right. \right. \notag \\
\left. \left. +8r^4a^4+10r^2a^6+15a^8 \right) \right]^{1/2} \mbox{.} 
\label{eq_k_12}  
\end{gather}

\section{Potentials, rotation curves and epicyclic frequencies for inverted spherical shells} \label{ap_D}
The expressions for the potentials $\Phi^{(m,n)}_{(inv.)}$ of the inverted shells (\ref{eq_shell_inv}) with $m=0,1$ 
and $n=1,2$ are
\begin{gather}
\Phi^{(0,1)}_{(inv.)}=-\frac{G\rho_c\pi a^3}{60r} \left[ \frac{15r}{a} \arcsin \left( \frac{a}{r} \right) +
\frac{\sqrt{r^2-a^2}}{r^5}\left( 17r^4+6a^2r^2-8a^4\right) \right] \mbox{,} \label{eq_phi_inv_01} \\
\Phi^{(0,2)}_{(inv.)}=-\frac{G\rho_c\pi a^3}{3360r} \left[ \frac{525r}{a} \arcsin \left( \frac{a}{r} \right)+
\frac{\sqrt{r^2-a^2}}{r^7}\left( 499r^6+162r^4a^2 \right. \right. \notag \\
\left. \left. +104r^2a^4-240a^6 \right) \right] \mbox{,} \\
\Phi^{(1,1)}_{(inv.)}=-\frac{G\rho_c\pi a^3}{1120r} \left[ \frac{105r}{a} \arcsin \left( \frac{a}{r} \right) +
\frac{\sqrt{r^2-a^2}}{r^7}\left( 151r^6+58r^4a^2 \right. \right. \notag \\
\left. \left. -184r^2a^4+80a^6 \right) \right] \mbox{,} \\
\Phi^{(1,2)}_{(inv.)}=-\frac{G\rho_c\pi a^3}{20160r} \left[ \frac{945r}{a} \left( \frac{a}{r} \right)+
\frac{\sqrt{r^2-a^2}}{r^9}\left( 1103r^8+394r^6a^2+264r^4a^4 \right. \right. \notag \\
\left. \left. -1712r^2a^6+896a^8 \right) \right] \mbox{.} \label{eq_phi_inv_12}  
\end{gather} 
The circular velocity $v_{c(inv.)}^{(m,n)}$ and epicyclic frequency $\kappa_{(inv.)}^{(m,n)}$ 
for  these inverted shells read as
\begin{gather}
 v_{c(inv.)}^{(0,1)}=\sqrt{\frac{4G\rho_c\pi a^3\left( 2r^2+3a^2\right)}{15r^3}} 
\left( 1- \frac{a^2}{r^2}\right)^{3/4} \mbox{,} \label{eq_vc01_inv} \\
v_{c(inv.)}^{(0,2)}=\sqrt{\frac{4G\rho_c\pi a^3\left( 8r^4+12r^2a^2+15a^4 \right)}{105r^5}}
\left( 1- \frac{a^2}{r^2}\right)^{3/4} \mbox{,} \label{eq_vc02_inv} \\
v_{c(inv.)}^{(1,1)}=\sqrt{\frac{4G\rho_c\pi a^3\left( 2r^2+5a^2\right)}{35r^3}}
\left( 1- \frac{a^2}{r^2}\right)^{5/4} \mbox{,} \label{eq_vc11_inv} \\
v_{c(inv.)}^{(1,2)}=\sqrt{\frac{4G\rho_c\pi a^3\left( 8r^4+20r^2a^2+35a^4 \right)}{315r^5}}
\left( 1- \frac{a^2}{r^2}\right)^{5/4} \mbox{,} \label{eq_vc12_inv} \\
\kappa_{(inv.)}^{(0,1)}=\sqrt{\frac{4G\rho_c\pi a^3\left( 2r^4+r^2a^2+12a^4\right)}{15r^7}}
\left( 1- \frac{a^2}{r^2}\right)^{1/4} \mbox{,} \label{eq_k01_inv} \\
\kappa_{(inv.)}^{(0,2)}=\sqrt{\frac{4G\rho_c\pi a^3\left( 8r^6+4r^4a^2+3r^2a^4+90a^6\right)}{105r^9}}
\left( 1- \frac{a^2}{r^2}\right)^{1/4} \mbox{,} \label{eq_k02_inv} \\
\kappa_{(inv.)}^{(1,1)}=\sqrt{\frac{4G\rho_c\pi a^3\left( 2r^6+r^4a^2+27r^2a^4-30a^6 \right)}{35r^9}}
\left( 1- \frac{a^2}{r^2}\right)^{1/4} \mbox{,} \\
\kappa_{(inv.)}^{(1,2)}=\sqrt{\frac{4G\rho_c\pi a^3\left( 8r^8+4r^6a^2+3r^4a^4+265r^2a^6-280a^8\right)}{315r^{11}}}
\left( 1- \frac{a^2}{r^2}\right)^{1/4} \mbox{.} \label{eq_k12_inv}
\end{gather}
\section{Potential, rotation curve and epicyclic frequency for a double shell} \label{ap_E}
The potential $\Phi^{(0,1)}_{(2)}$ of the double shell discussed in Section \ref{sec_sph_d_sh2} 
reads as
\begin{multline}
\Phi^{(0,1)}_{(2)}=-\frac{G\rho_c\pi}{20160a^7} \left\{ \frac{315a^{10}}{r} 
\left( 7b^4-20b^2+16 \right) \arcsin \left( \frac{r}{a} \right) \right. \\
\left. + \sqrt{a^2-r^2} \left[ -896r^8b^4+16r^6a^2b^2 \left( 17b^2+180 \right) 
+24r^4a^4 \left( 15b^4-52b^2-112 \right) \right. \right. \\
\left. \left. + 2r^2a^6 \left( 289b^4-972b^2+1008 \right) 
+ a^8 \left( 1891b^4-5988b^2+5712 \right) \right] \right\} \mbox{.} 
\label{eq_phi_2_01}
\end{multline}
The corresponding expressions of the circular velocity $v_{c(2)}^{(0,1)}$ 
and epicyclic frequency $\kappa_{(2)}^{(0,1)}$ are
\begin{gather}
v_{c(2)}^{(0,1)}=\sqrt{\frac{G\rho_c\pi}{960a^7}} \left\{ \frac{15a^{10}}{r}
\left( 7b^4-20b^2+16 \right) \arcsin \left( \frac{r}{a} \right) \right. \notag \\
\left. + \sqrt{a^2-r^2} \left[ 384r^8b^4-48r^6a^2b^2 \left( b^2+20 \right) 
\ -8r^4a^4 \left( 7b^4-20b^2-80 \right) \right. \right. \notag \\
\left. \left. -10r^2a^6 \left( 7b^4-20b^2+16 \right) 
 - 15a^8 \left( 7b^4-20b^2+16 \right) \right] \right\}^{1/2}
\mbox{,} \label{eq_vc_2_01} \\
\kappa_{(2)}^{(0,1)}=\sqrt{\frac{G\rho_c\pi}{960r^2a^7}} \left\{ \frac{15a^{10}}{r}
\left( 7b^4-20b^2+16 \right) \arcsin \left( \frac{r}{a} \right) \right. \notag \\
\left. + \sqrt{a^2-r^2} \left[ 4224r^8b^4-48r^6a^2b^2 \left( b^2+180 \right)  
-8r^4a^4 \left( 7b^4-20b^2-560 \right) \right. \right. \notag \\
\left. \left. -10r^2a^6 \left( 7b^4-20b^2+16 \right)
 - 15a^8 \left( 7b^4-20b^2+16 \right) \right] \right\}^{1/2}
\mbox{.} \label{eq_k_2_01}
\end{gather}

\begin{thebibliography}{99}
\bibitem{pl11} Plummer H. C., 1911, MNRAS, 71, 460
\bibitem{ja83} Jaffe W., 1983, MNRAS, 202, 995
\bibitem{her90} Hernquist L., 1990, ApJ, 356, 359
\bibitem{nav97} Navarro J. F., Frenk C. S., White S. D. M., 1997, ApJ, 490, 493
\bibitem{de93} Dehnen W., 1993, MNRAS, 265, 250
\bibitem{tr94} Tremaine S., Richstone D. O.,
Byun Y. I., Dressler A., Faber S. M., Grillmair C., Kormendy J., Lauer T. R., 1994, AJ, 107, 634
\bibitem{zh96} Zhao H. S., 1996, MNRAS, 278, 488
\bibitem{bs99} Bi\v{c}\'{a}k J., Schmidt B. G., 1999, ApJ, 521, 708
\bibitem{re99} Rein G., 1999, Indiana Univ. Math. J., 48, 335
\bibitem{an07a} Andr\'easson H., 2007, Communications Math. Phys., 274, 409
\bibitem{an07b} Andr\'easson H., Rein G., 2007, Classical Quantum Gravity, 24, 1809
\bibitem{gr10} Gleiser R. J., Ramirez M. A., 2010, Classical Quantum Gravity, 27, 065008
\bibitem{vl10} Vogt D., Letelier P. S., 2010, MNRAS, 402, 1313
\bibitem{kal72} Kalnajs A. J., 1972, ApJ, 175, 63
\bibitem{gon06} Gonz\'alez G. A., Reina J. I., 2006, MNRAS, 371, 1873
\bibitem{mor69} Morgan T., Morgan L., 1969, Phys. Rev., 183, 1097
\bibitem{bt08} Binney J., Tremaine S., 2008,
Galactic Dynamics, 2nd edn. Princeton Univ. Press, Princeton, NJ
\bibitem{kel47} Thomson W. (Lord Kelvin), 1847, J. Math. Pures Appliquees, 12, 256
\bibitem{kel53} Kellog O. D., 1953, Foundations of Potential Theory, Dover Publications, New York 
\bibitem{let07} Letelier P. S., 2007, MNRAS, 381, 1031
\end{thebibliography}
\end{document}